\documentclass[12pt]{JHEP3}

\usepackage{amsmath,epsfig}
\usepackage{amssymb,amsfonts}
\usepackage{latexsym}
\usepackage{longtable}

\relax

\def\be{\begin{equation}}
\def\ee{\end{equation}}
\def\bea{\begin{eqnarray}}
\def\eea{\end{eqnarray}}

\newcommand\fverb{\setbox\pippobox=\hbox\bgroup\verb}
\newcommand\fverbdo{\egroup\medskip\noindent%
                        \fbox{\unhbox\pippobox}\ }
\newcommand\fverbit{\egroup\item[\fbox{\unhbox\pippobox}]}

\newcommand{\bear}{\begin{eqnarray}}

\newcommand{\eear}{\end{eqnarray}}

\newbox\pippobox

\def\6{\partial}

\def\a{\alpha}

\def\sq
\def\a{\alpha}

\def\bx{{\bf x}}

\def\beq{\begin{equation}}
\def\eeq{\end{equation}}
\def\bea{\begin{eqnarray}}
\def\eea{\end{eqnarray}}
\def\bean{\begin{eqnarray*}}
\def\eean{\end{eqnarray*}}
\def\bx{{\bf x}}
\def\bk{{\bf k}}

\title{Perturbative instabilities in Ho\v{r}ava gravity}

\author{{\large Charalampos Bogdanos }
~\\
~\\
LPT, Universit$\acute{e}$ de Paris-Sud-11, B$\hat{a}$t. 210, 91405
Orsay CEDEX, France {\tt Charalampos.Bogdanos@th.u-psud.fr}}

\author{{\large Emmanuel N. Saridakis }
~\\
~\\
Department of Physics, University of Athens, GR-15771 Athens,
Greece {\tt msaridak@phys.uoa.gr}}

\preprint{}

\abstract{We investigate the scalar and tensor perturbations in
Ho\v rava gravity, with and without detailed balance, around a
flat background. Once both types of perturbations are taken into
account, it is revealed that the theory is plagued by ghost-like
scalar instabilities in the range of parameters which would render
it power-counting renormalizable, that cannot be overcome by
simple tricks such as analytic continuation. Implementing a
consistent flow between the UV and IR limits seems thus more
challenging than initially presumed, regardless of whether the
theory approaches General Relativity at low energies or not. Even
in the phenomenologically viable parameter space, the tensor
sector leads to additional potential problems, such as
fine-tunings and super-luminal propagation.}

\begin{document}

\section{Introduction}

Relativity is commonly assumed to be the foundation of every
gravitational construction, especially in its cosmological
application. However, since a quantum field theory of general
relativity proves to be non-renormalizable, numerous attempts of
its modification have appeared in the literature. Recently, a
power-counting renormalizable, ultra-violet (UV) complete theory
of gravity was proposed by Ho\v{r}ava in
\cite{hor2,hor1,hor3,hor4}. Although presenting General Relativity
as an infrared (IR) fixed point, in the UV the theory possesses a
fixed point with an anisotropic, Lifshitz scaling between time and
space.

Due to these novel features, there has been a large amount of
effort in examining and extending the properties of the theory
itself
\cite{Volovik:2009av,Cai:2009ar,Cai:2009dx,Orlando:2009en,Nishioka:2009iq,Konoplya:2009ig,Charmousis:2009tc,Li:2009bg,Visser:2009fg,Sotiriou:2009gy,Sotiriou:2009gybb,Chen:2009bu,Chen:2009ka,Shu:2009gc,Dutta:2009jn,Dutta:2010jh}.
Additionally, application of Ho\v{r}ava gravity as a cosmological
framework gives rise to Ho\v{r}ava cosmology, which proves to lead
to interesting behavior \cite{Calcagni:2009ar,Kiritsis:2009sh}. In
particular, one can examine specific solution subclasses
\cite{Lu:2009em,Nastase:2009nk,Colgain:2009fe,Ghodsi:2009rv,Minamitsuji:2009ii,Ghodsi:2009zi},
the perturbation spectrum
\cite{Mukohyama:2009gg,Piao:2009ax,Gao:2009bx,Chen:2009jr,Gao:2009ht},
the gravitational wave production
\cite{Mukohyama:2009zs,Takahashi:2009wc,Koh:2009cy}, the matter
bounce \cite{Brandenberger:2009yt,Brandenberger:2009ic}, the black
hole properties
\cite{Danielsson:2009gi,Cai:2009pe,Myung:2009dc,Kehagias:2009is,Cai:2009qs,Mann:2009yx,Bertoldi:2009vn,Castillo:2009ci,BottaCantcheff:2009mp},
the dark energy phenomenology
\cite{Saridakis:2009bv,Park:2009zr,Wang:2009rw,Leon:2009rc} etc.

However, investigations of various aspects of Ho\v{r}ava gravity
have started to reveal some potentially troublesome features.
Although some of them can be alleviated or removed once the
detailed-balance condition is relaxed
\cite{Li:2009bg,Mukohyama:2009mz,Mukohyama:2009tp}, the most
significant core remains unaffected. In \cite{Charmousis:2009tc}
it was shown that the diffeomorphism-invariance breaking leads to
an additional degree of freedom (known already in \cite{hor1} as
the longitudinal degree of freedom of metric perturbations), which
is not needed to decouple in the IR, and thus General Relativity
is not recovered at any scale. Similarly, in \cite{Blas:2009yd} it
was shown that the explicit breaking of general covariance
uncovers an extra scalar degree of freedom, with fast exponential
instabilities at short distances and strong-coupling at extremely
low cutoff scales. In addition, observational constraints may rule
out the theory completely \cite{Calcagni:2009qw}. These features
led many authors to examine various forms of ``modified''
Ho\v{r}ava gravity, such as versions with full diffeomorphism
invariance \cite{Germani:2009yt,Nojiri:2009th}, with deformed
action and zero cosmological constant (and thus with Minkowski and
not AdS IR limit) \cite{Myung:2009va,myangbb}, with restored
parity invariance beyond detailed-balance
\cite{Visser:2009fg,Sotiriou:2009gy,Sotiriou:2009gybb}, with extra
derivative-terms \cite{Cai:2009in}, and with ``soft''
detailed-balance breaking \cite{Park:2009zra}.

On the other hand, most works on the cosmological application of
Ho\v{r}ava gravity focus straightaway on its IR limit, as it is
realistically  expected to happen, without examining the running
of the theory towards this regime. Indeed, in the original papers
of Ho\v{r}ava the running behavior was revealed, but not the
precise flow. It is interesting to point out that, up to this
point, there is an amount of ambiguity concerning the IR behavior
of the theory. The presence of additional degrees of freedom
appears to be in some sense background-dependent. Although strong
coupling arises once the perturbations around a flat Minkowski
spacetime are considered \cite{Charmousis:2009tc}, the theory
seems to be well behaved when perturbing a cosmological background
\cite{Gao:2009ht}. It is therefore non-trivial to trace internal
consistency problems of the theory directly at the IR limit, at
least in the perturbative level, even though it is already known
that non-perturbative solutions such as the Schwarzschild black
hole are not recovered in the context of the conventional,
detailed-balance-preserving Ho\v{r}ava gravity \cite{Lu:2009em}.

In the present work we are interested in performing a detailed
investigation of the gravitational perturbations of Ho\v{r}ava
gravity, using it as a tool to examine its consistency. We study
both scalar and tensor sectors, around a Minkowski background. We
find that instabilities do appear in the whole range of parameter
values, between which the theory is supposed to flow in order to
provide a consistent UV completion. Such instabilities seem to
exclude any viable flow of the theory within the range needed to
provide power-counting renormalizability. Breaking detailed
balance also does not seem to improve the situation. Additionally,
these results are not affected by the above mentioned ambiguity
about the existence or not of additional degrees of freedom at the
IR. Finally, we argue that the IR limit behavior may actually be
the least of the problems.

In summary, the gravitational perturbation investigation reveals
that Ho\v{r}ava gravity, in its present form, suffers from
instabilities and fine-tunings that seem to originate from its
deep and implicit features and thus they cannot be overcome by
simple tricks such as analytic continuation. The paper is
organized as follows: In section \ref{perturbations} we extract
the scalar and tensor perturbations under the ADM diffeomorphism
group in Minkowski spacetime, setting the relevant degrees of
freedom and fixing the gauge. In section \ref{discussion} we
discuss about the emerging instabilities, the fine-tunings and the
possible causality problems. Finally, section \ref{Conclusions} is
devoted to a summary and discussion of the obtained results.

\section{Gravitational perturbations in Minkowski background}
\label{perturbations}

One of the most decisive tests for the reliability of a
gravitational theory is the examination of its perturbations. This
investigation reveals some of the deep features of the theory, and
additionally, one may hope to eventually extract possible
observational signatures. Thus, the study of gravitational
perturbations of Ho\v{r}ava gravity will be the basic tool of the
present work.

\subsection{Variables and gauge transformations}
\label{gauge}

We consider coordinate transformations of the form $x^{\mu} \to
\tilde x^{\mu}=x^{\mu}+\xi^{\mu}$. Under this transformation the
metric-perturbation around a given background changes as $\delta
\tilde g_{\mu \nu} = \delta g_{\mu \nu}-\nabla_{\mu}
\xi_{\nu}-\nabla_{\nu} \xi_{\mu}$.

As it is known, the perturbation analysis, especially its
primordial phase, is sensitive to the background evolution. A
usual choice for cosmological applications
\cite{Calcagni:2009ar,Kiritsis:2009sh}, is to impose $N^{(0)}=1$,
$N_{i}^{(0)}=0$, $g_{ij}^{(0)}=a^2(t)\delta _{ij}$ in the ADM
(``foliation-preserving") formalism. Although we will soon perform
perturbations around  a flat background, we keep for the moment
the explicit dependence on the scale factor for the sake of
generality. The background metric is given by
\begin{eqnarray}
ds^2_{(0)}&=&(g^{(0)}_{ij}N^{i(0)}N^{j(0)}-N^2)dt^2+2g^{(0)}_{ij}N^{i(0)} dt dx^j+g^{(0)}_{ij}dx^i dx^j \nonumber \\
&=&a^2(n) \left( -dn^2+\delta_{ij} dx^i dx^j \right).
\end{eqnarray}
From now on the $0$-coordinate  denotes the conformal time $n$,
defined through $dt^2/a^2=dn^2$. Therefore, the general
perturbations of the metric around this background read:
\begin{eqnarray}
\delta g_{00}&=&-2a^2\phi\\
\delta g_{0i}&=&a^2\partial_i B+a^2Q_i\\
\delta g_{ij}&=&a^2h_{ij}-a^2(\partial_i W_j+\partial_j W_i)-2a^2
\psi \delta_{ij}+2 a^2 \partial_i \partial_j E.
 \end{eqnarray}
The vector modes are assumed to be transverse, that is $\partial_i
W^i =\partial_i Q^i=0$, while  the tensor mode is forced to be
transverse and traceless, $\partial_i h^{ij}=\delta^{ij}h_{ij}=0$.
Finally, the Christoffel symbols read
\begin{equation}
 \Gamma^0_{00}=H, \,\, \Gamma^i_{0j}=H \delta^i_j,
\,\, \Gamma^0_{ij}=H \delta_{ij},
 \end{equation}
where $H$ is the {\it conformal Hubble parameter},
$H=\frac{1}{a(n)} \frac{da(n)}{dn}$. In the following, we
concentrate on the flat geometry, and since there is now no
distinction between coordinate and conformal time we restore $t$
as denoting the time coordinate.

The spatial part of the gauge transformation vector is defined as
$\xi^i=\partial^i \xi+\zeta^i$ with $\zeta_i=\zeta^i$, so that
$\xi_i=a^2(\partial_i \xi+\zeta_i)$. Anticipating the application
to the Ho\v{r}ava gravity framework, where only transformations of
the time coordinate of this sort are allowed, we assume that
$\xi^0=\xi^0(t)$. Using these transformation rules, we obtain the
following expressions for the gauge transformation of each mode
(note that $\xi_0=-a^2\xi^0$):
\begin{itemize}
 \item Scalars
\begin{eqnarray}
\tilde \psi &=& \psi +H\xi^0\\
\tilde \phi &=& \phi -H\xi^0 -\dot \xi^0 \label{phigauge}\\
\tilde B &=& B - \dot \xi\label{Bgauge}\\
\tilde E &=& E - \xi
\end{eqnarray}
\item Vectors
\begin{eqnarray}
\tilde Q_i= Q_i -\dot \zeta_i\label{Qgauge}\\
\tilde W_i= W_i + \zeta_i
\end{eqnarray}
\item Tensor
\begin{equation}
\tilde h_{ij} = h_{ij},
\end{equation}
\end{itemize}
where a dot denotes the coordinate time-derivative. Lastly, note
that since we restrict ourselves to the flat case, in the
following we set $H=0$.

Let us now discuss about the gauge fixing, which is required for
the action derivation and the determination of the physical
degrees of freedom. The projectability condition of Ho\v{r}ava
gravity \cite{hor3} requires that the perturbation of the
lapse-function $N$ depends only on time, thus $\phi\equiv\phi(t)$.
This allows us to ``gauge away'' the $\phi$-perturbation, that is
to set $\tilde \phi=0$ imposing $\phi=\dot \xi^0$. Similarly, we
fix $\tilde B=0$ from setting $B=\dot \xi$. Finally, setting
$Q_i=\dot \zeta_i$ we eliminate the $Q_i$ degree of freedom.
Therefore, the remaining degrees of freedom are $\psi$, $E$, $W_i$
and $h_{ij}$.

In summary, in the aforementioned gauge we obtain
\begin{eqnarray}
\delta N &=& \delta N_i =0\\
\delta_{ij} &=& h_{ij}-2\psi\delta_{ij}+2\partial_i\partial_j
E-(\partial_i W_j+\partial_j W_i).
 \end{eqnarray}
Note that since only perturbations imposed on the ``same-time''
spatial hypersurface are allowed, this is equivalent to a {\it
synchronous gauge} choice.

\subsection{Perturbed Action}

In Ho\v{r}ava gravity the gravitational action can be decomposed
into a kinetic and a potential part as $S_g = S_K+S_V$, where
\begin{equation}
\label{kinetic}
 S_K=\frac{2}{\kappa^2}\int dtd^3x
\sqrt{g}N\left(K_{ij}K^{ij}-\lambda K^2\right),
\end{equation}
with
\begin{equation}\label{extri-curv} K_{ij}=\frac{1}{2N}(\dot
g_{ij}-\nabla_{i}N_j-\nabla_jN_i)\;,
\end{equation}
 the extrinsic curvature and $\kappa$ a constant
with mass dimension $-1$.

One of the subjects that have led to discussion in the literature
is the imposition of detailed balance \cite{hor3}, which apart
from reducing the possible terms in the action, it allows for a
quantum inheritance principle \cite{hor2} (the $D+1$ dimensional
theory acquires the renormalization properties of the
D-dimensional one). In order to proceed and write explicitly the
potential term for the moment we impose detailed balance too.
However, in order to avoid possible accidental artifacts of this,
possibly ambiguous, condition in the obtained results, later on we
will extend our analysis beyond detailed balance.

Under detailed balance we can write
\begin{eqnarray}
\label{potential} S_{V}=\int dt d^3x\sqrt{g}N
\left[-\frac{\kappa^2}{2w^4}C_{ij}C^{ij}\right.
&+&\frac{\kappa^2\mu}{2w^2}\frac{\epsilon^{ijk}}{\sqrt{g}}R_{il}\nabla_j R^l_k-\frac{\kappa^2\mu^2}{8}R_{ij}R^{ij} \nonumber\\
&+&\left.
\frac{\kappa^2\mu^2}{8(1-3\lambda)}\left(\frac{1-4\lambda}{4}R^2
+\Lambda R - 3\Lambda^2 \right) \right].\ \ \ \ \ \
\end{eqnarray}
In the above expression, the Cotton tensor of the spatial
hypersurface is given by
 \begin{equation}
C^{ij}=\frac{\epsilon^{ikl}}{\sqrt{g}}\nabla_k \left(
R^j_l-\frac{1}{4} R \delta^j_l \right),
 \end{equation}
 and the covariant derivatives are defined with
respect to the spatial metric $g_{ij}$. $\epsilon^{ijk}$ is the
totally antisymmetric unit tensor, $\Lambda $ is a negative
constant which is related to the cosmological constant in the IR
limit, and the variables $w$ and $\mu$ are constants with mass
dimensions $0$ and $1$, respectively. Finally, $\lambda$ is a
dimensionless constant, which incorporates the running behavior of
Ho\v{r}ava gravity.  Although a priori arbitrary, it is now known
that the important variation-regime is between $1/3$ and $1$, with
the former corresponding to the UV limit and the later to the IR
one.

We will now perturb the Ho\v{r}ava gravitational action up to
second order. At this point, we should mention that varying the action once we have substituted our metric ansatz for the perturbations is equivalent to deriving the Hamilton and momentum constraints along with the remaining equations of motion for an arbitrary metric perturbation and then imposing the ansatz. We choose the former approach here for reasons of simplicity.  After non-trivial but straightforward calculations
that are presented in Appendix \ref{appendix}, for the
perturbed kinetic part (\ref{kinetic}) we obtain
\begin{equation}
\delta S^{(2)}_K=\int dt d^3x \frac{2}{\kappa^2}\left[\frac{1}{4}
\dot h_{ij} \dot h^{ij} +(1-3 \lambda) \left( 3 \dot \psi^2 - 2
\dot \psi \nabla^2 \dot E \right) + (1-\lambda) \dot E \nabla^4
\dot E \right],
 \label{DSK}
  \end{equation}
  while for the perturbed potential part
(\ref{potential}) we acquire
\begin{eqnarray} \delta S_V^{(2)}  &=& \int d
td^3 x\left[ {\frac{{\kappa ^2 }} {{8w^4 }}h_{ij} \nabla ^6 h^{ij}
+ \frac{{\kappa ^2 \mu }} {{8w^2 }}\epsilon^{ijk} h_{il} \partial
_j \nabla ^4 h_k^l  - \frac{{\kappa ^2 \mu ^2 }}
{{32}}h_{ij} \nabla ^4 h^{ij} } \right.\: \nonumber \\
&& \;\;\;\;\;\;\;\;\;\;\;\;\;\;\;+ \frac{{\kappa ^2 \mu ^2 \Lambda
}} {{32(1 - 3\lambda )}}h_{ij} \nabla ^2 h^{ij}  - \frac{{\kappa
^2 \mu ^2 (1 - \lambda )}} {{4(1 - 3\lambda )}}\psi \nabla ^4 \psi
- \frac{{\kappa ^2 \mu ^2 \Lambda }}
{{4(1 - 3\lambda )}}\psi \nabla ^2 \psi  \nonumber \\
&& \;\;\;\;\;\;\;\;\;\;\;\;\;\;\; \left. { + \frac{{27\kappa ^2
\mu ^2 \Lambda ^2 }} {{16(1 - 3\lambda )}}\psi ^2  -
\frac{{9\kappa ^2 \mu ^2 \Lambda ^2 }} {{8(1 - 3\lambda )}}\psi
\nabla ^2 E + \frac{{3\kappa ^2 \mu ^2 \Lambda ^2 }} {{16(1 -
3\lambda )}}E\nabla ^4 E} \right] \label{DSV}.
 \end{eqnarray}

\subsection{Scalar perturbations}

As can be observed from (\ref{DSK}),(\ref{DSV}) the action for
scalar perturbations includes the two modes $\psi$ and $E$ and it
is written as
\begin{eqnarray}
\delta S_S^{(2)}  &=& \int d td^3
x\left[ {\frac{{2(1 - 3\lambda )}} {{\kappa ^2 }}\left( {3\dot
\psi ^2  - 2\dot \psi \nabla ^2 \dot E} \right) + \frac{{2(1 -
\lambda )}}
{{\kappa ^2 }}\dot E\nabla ^4 \dot E} \right. \nonumber\\
&&\;\;\;\;\;\;\;\;\;\;\;\;\;\;\; - \frac{{\kappa ^2 \mu ^2 (1 -
\lambda )}} {{4(1 - 3\lambda )}}\psi \nabla ^4 \psi  -
\frac{{\kappa ^2 \mu ^2 \Lambda }} {{4(1 - 3\lambda )}}\psi \nabla
^2 \psi  + \frac{{27\kappa ^2 \mu ^2 \Lambda ^2 }}
{{16(1 - 3\lambda )}}\psi ^2 \nonumber \\
&&\;\;\;\;\;\;\;\;\;\;\;\;\;\;\;\left. { - \frac{{9\kappa ^2 \mu
^2 \Lambda ^2 }} {{8(1 - 3\lambda )}}\psi \nabla ^2 E +
\frac{{3\kappa ^2 \mu ^2 \Lambda ^2 }} {{16(1 - 3\lambda
)}}E\nabla ^4 E} \right].
 \end{eqnarray}
 Varying it with respect to $E$ and $\psi$ we obtain the equations of motion:
 \begin{equation}
  \frac{8} {{\kappa ^2
}}\ddot E + \frac{{\kappa ^2 \mu ^2 \left( {1 - \lambda }
\right)}} {{2\left( {1 - 3\lambda } \right)}}\nabla ^2 \psi  +
\frac{{\kappa ^2 \mu ^2 \Lambda }} {{2\left( {1 - 3\lambda }
\right)}}\psi  = 0\,
 \label{alphaeq}
 \end{equation}
\begin{eqnarray}
 \frac{8} {{\kappa ^2 }}\frac{{1 - 3\lambda }} {{1
- \lambda }}\ddot \psi &-& \frac{{9\kappa ^2 \mu ^2 \Lambda ^2 }}
{{4\left( {1 - \lambda } \right)\left( {1 - 3\lambda }
\right)}}\psi  + \frac{{3\kappa ^2 \mu ^2 \Lambda ^2 }} {{4\left(
{1 - \lambda } \right)\left( {1 - 3\lambda } \right)}}\nabla ^2 E
\nonumber \\  &+& \frac{{\kappa ^2 \mu ^2 \left( {1 - \lambda }
\right)}} {{2\left( {1 - 3\lambda } \right)}}\nabla ^4 \psi+
\frac{{\kappa ^2 \mu ^2 \Lambda }} {{2\left( {1 - 3\lambda }
\right)}}\nabla ^2 \psi  = 0.
 \label{betaeq}
 \end{eqnarray}
As can be seen these two equations are coupled, not allowing for a
straightforward stability investigation. However, we can still
acquire information about the stability of the configuration by
studying it at high and low momenta. Taking the IR limit of
(\ref{alphaeq}), (\ref{betaeq}), that is considering their low-$k$
behavior, they reduce to
 \begin{equation}
 \label{psi00}
 \frac{8} {{\kappa ^2
}}\ddot E + \frac{{\kappa ^2 \mu ^2 \Lambda }} {{2\left( {1 -
3\lambda } \right)}}\psi  = 0
\end{equation}
 \begin{equation}
  \frac{8} {{\kappa ^2
}}\frac{{1 - 3\lambda }} {{1 - \lambda }}\ddot \psi  -
\frac{{9\kappa ^2 \mu ^2 \Lambda ^2 }} {{4\left( {1 - \lambda }
\right)\left( {1 - 3\lambda } \right)}}\psi  = 0
\label{psi1}.
\end{equation}
 Thus, the second equation is decoupled, acting as a low-momentum
equation of motion for the scalar field $\psi$.

A straightforward observation from (\ref{psi1}) is that it leads
to a ghost-like behavior in the IR limit whenever $\frac{1}{3} <
\lambda < 1$. Inverting the overall sign of the Lagrangian is not
going to help, since as we will promptly see, the time derivative
of the tensor perturbation has the opposite sign. In particular,
(\ref{psi1})  leads to
 \begin{equation}
  \ddot {\tilde  \psi}  - \frac{{9\kappa ^4 \mu ^2 \Lambda ^2 }} {{32\left(
{1 - 3\lambda } \right)^2 }} \tilde \psi = 0,
\end{equation}
where
\begin{eqnarray}
\psi(t,\bx)=\int \frac{d^3k}{(2\pi)^\frac{3}{2}}
~\tilde{\psi}_k(t)e^{i\bk\cdot\bx}.
\end{eqnarray}
Therefore, we acquire the following dispersion relation
 \begin{equation}
\omega^2\equiv m^2 = - \frac{{9\kappa ^4 \mu ^2 \Lambda ^2 }}
{{32\left( {1 - 3\lambda } \right)^2 }} < 0
\label{masseq},
 \end{equation}
which induces instabilities at the IR, regardless of the
$\lambda$-value
 and of the sign of the cosmological constant. Finally,
combination of both (\ref{psi00}),(\ref{psi1}) fixes $E$ against
$\psi$, and in particular it leads to
 \begin{equation}
\psi=-\frac{9}{2}\frac{\Lambda}{1-3\lambda} E.
 \label{psiE}
 \end{equation}
Thus, we are led to only one effective degree of freedom, or in
other words the system has been diagonalized.

Now, for high $k$,  (\ref{alphaeq}), (\ref{betaeq}) reduce to
\begin{equation}
 \frac{8} {{\kappa ^2 }}\ddot E + \frac{{\kappa ^2
\mu ^2 \left( {1 - \lambda } \right)}} {{2\left( {1 - 3\lambda }
\right)}}\nabla ^2 \psi  = 0
 \end{equation}
  \begin{equation}
\frac{8} {{\kappa ^2 }}\frac{{1 - 3\lambda }} {{1 - \lambda
}}\ddot \psi  + \frac{{\kappa ^2 \mu ^2 \left( {1 - \lambda }
\right)}} {{2\left( {1 - 3\lambda } \right)}}\nabla ^4 \psi  = 0.
\label{psi2}
 \end{equation}
Therefore, the ghost-like coefficient in the time derivative part
remains and  (\ref{psi2}) yields a high-$k$ dispersion relation of
the form
 \begin{equation}
   \omega ^2  \equiv
  \frac{{\kappa ^4 \mu ^2 }}
{{16}}\left( {\frac{{1 - \lambda }} {{1 - 3\lambda }}} \right)^2
k^4.
 \label{omegascalar}
  \end{equation}

In general the homogeneous system of equations
(\ref{alphaeq}),(\ref{betaeq}) leads to only one active degree of
freedom (one of the fields should be defined in terms of the
other, in order to have a non-trivial solution) and a dispersion
relation of the form
\begin{eqnarray}
 \frac{{64}} {{k^4
}}\left(\frac{{1 - 3\lambda }} {{1 - \lambda }}\right)\omega ^4
&+& \left[ {\frac{{18\mu ^2 \Lambda ^2 }} {{\left( {1 - \lambda }
\right)\left( {1 - 3\lambda } \right)}} + \frac{{4\mu ^2 \Lambda
}} {{1 - 3\lambda }}\,k^2  - \frac{{4\mu ^2 \left( {1 - \lambda }
\right)}}
{{1 - 3\lambda }}\,k^4 } \right]\omega ^2 \nonumber \\
 &+& \frac{{3\kappa ^4 \mu ^4 \Lambda ^2 }}
{{8\left( {1 - \lambda } \right)\left( {1 - 3\lambda } \right)^2
}}\left[{\Lambda  + \left( {\lambda  - 1} \right)k^2 } \right]k^2
= 0.
\end{eqnarray}
Therefore, the physical requirement of obtaining a positive
solution for $\omega^{2}$ will lead to restrictions on the various
parameters of the theory.

\subsection{Tensor perturbations}

Let us now examine the tensor perturbations. Their action can be
extracted from (\ref{DSK}),(\ref{DSV}) and it reads:
\begin{eqnarray}
\delta S^{(2)}_T &=& \int dt d^3x \left[ \frac{1}{2\kappa^2} \dot
h_{ij} \dot h^{ij}
 + \frac{{\kappa ^2 \mu ^2 \Lambda }}
{{32(1 - 3\lambda )}}h_{ij} \nabla ^2 h^{ij} \right. \nonumber \\
&&\;\;\;\;\;\;\;\;\;\;\;\;\;\;\;  \left.+ \frac{\kappa^2}{8w^4}
h_{ij} \nabla^6 h^{ij} + \frac{\kappa^2 \mu}{8 w^2} \epsilon^{ijk}
h_{il} \partial_j \nabla^4 h^l_k - \frac{\kappa^2 \mu^2}{32}
h_{ij} \nabla^4 h^{ij}\right].
\end{eqnarray}
Therefore, the graviton equation of motion writes as
\begin{equation}
\ddot h^{ij}- \frac{{\kappa ^4 \mu ^2 \Lambda }} {{16(1 - 3\lambda
)}}\nabla ^2 h^{ij}  - \frac{\kappa^4}{4 w^4} \nabla^6 h^{ij} -
\frac{\kappa^4 \mu}{4 w^2} \epsilon^{ilk} \partial_l \nabla^4
h^j_k + \frac{\kappa^4 \mu^2}{16} \nabla^4 h^{ij} = 0.
\end{equation}
 Assuming graviton propagation along the $x^3$
direction, that is $k_i=k^i=(0,0,k)$, the $h_{ij}$ can be written
as usual in terms of polarization components as
\begin{equation}
h_{ij}=h^{ij}=\left(
\begin{array}{ccc}
h_+\,&\,h_{\times}\,&\,0\\
h_{\times}\,&\,-h_+&\,0\\
0\,&\,0\,&\,0
\end{array} \right).
\end{equation}
Using this parametrization we derive the two equations for the
different polarizations
\begin{equation}
\label{eomplus}
 \ddot h_+ - \frac{\kappa^4}{4 w^4} \nabla^6
h_+
 + \frac{\kappa^4 \mu}{4 w^2} \partial_3 \nabla^4 h_{\times} + \frac{\kappa^4 \mu^2}{16} \nabla^4 h_+ -
 \frac{{\kappa ^4 \mu ^2 \Lambda }}
{{16(1 - 3\lambda )}}\nabla ^2 h_{+}  = 0,
\end{equation}
\begin{equation}
\label{eomtimes} \ddot h_{\times} - \frac{\kappa^4}{4 w^4}
\nabla^6 h_{\times} -
 \frac{\kappa^4 \mu}{4 w^2} \partial_3 \nabla^4 h_+ + \frac{\kappa^4 \mu^2}{16} \nabla^4 h_{\times} -
 \frac{{\kappa ^4 \mu ^2 \Lambda }}
{{16(1 - 3\lambda )}}\nabla ^2 h_{\times}  = 0,
\end{equation}
and thus we identify the light speed from the low $k$ regime as
\begin{equation}
\label{lightspeed} c^2  = \frac{{\kappa ^4 \mu ^2 \Lambda }}
{{16(1 - 3\lambda )}}.
\end{equation}
A significant observation is that the two polarization modes are
coupled due to the $\epsilon^{ijk}$ term, which arises from the
Cotton tensor. Thus, it is more convenient to shift to the
Left-Right base defining
\begin{eqnarray}
h_L \equiv \frac{1}{\sqrt{2}} (h_+ + ih_\times),\\
h_R \equiv \frac{1}{\sqrt{2}} (h_+ - ih_\times).
\end{eqnarray}
In this case, and after Fourier transforming we obtain the
decoupled equations of motion (\ref{eomplus}),(\ref{eomtimes}):
\begin{eqnarray}
-\omega^2\tilde{h}_L + c^2k^2\tilde{h}_L +
\frac{\kappa^4\mu^2}{16}k^4\tilde{h}_L +
\frac{\kappa^4\mu}{4w^2}k^5\tilde{h}_L
+ \frac{\kappa^4}{4w^4}k^6\tilde{h}_L = 0,\\
-\omega^2\tilde{h}_R + c^2k^2\tilde{h}_R +
\frac{\kappa^4\mu^2}{16}k^4\tilde{h}_R -
\frac{\kappa^4\mu}{4w^2}k^5\tilde{h}_R +
\frac{\kappa^4}{4w^4}k^6\tilde{h}_R = 0,
\end{eqnarray}
an equation system that accepts a non-trivial solution only if the
corresponding determinant is zero, which leads to the dispersion
relation
\begin{equation}
\label{omegatensor}
 \omega ^2  = c^2 k^2  + \frac{{\kappa ^4 \mu ^2
}} {{16}}k^4  \pm \frac{{\kappa ^4 \mu }} {{4w^2 }}k^5  +
\frac{{\kappa ^4 }} {{4w^4 }}k^6.
\end{equation}
Therefore, we can identify the effective light speed as
\begin{eqnarray}
\label{speedtensor} c_{s}^2(k) &=& c^2   + \frac{{\kappa ^4 \mu ^2
}} {{16}}k^2 \pm \frac{{\kappa ^4 \mu }} {{4w^2 }}k^3  +
\frac{{\kappa ^4 }} {{4w^4
}}k^4\nonumber\\
&=&c^2\left[1+\frac{(1-3\lambda)}{\Lambda}k^2\left(1\pm\frac{2}{w^2\mu}k\right)^2\right].
\end{eqnarray}
In expressions (\ref{omegatensor}),(\ref{speedtensor}) the plus
and minus branches correspond to Left-handed and Right-handed mode
respectively.

Now, since the signs of the $k^5$-term are different in the
Left-handed and Right-handed modes, they propagate with a
different speed, leading to a rotation of the polarization plane,
and the rotation angle can be calculated by
\begin{eqnarray}
\delta\chi = \int_i^f (\omega_R-\omega_L) dt,
\end{eqnarray}
where $\omega_R$ and $ \omega_L$ represent the frequency of the
Right-handed and Left-handed mode respectively and the subscripts
"i" and "f" denotes the initial and final moment of the
propagation. Note that due to this rotation, the spectrum of
tensor perturbations would be expected to be suppressed at high
energy scales, as was shown in a different context in
\cite{Cai:2007xr}.

\subsection{Beyond Detailed Balance}
\label{nodetbal}

In the above analysis we were restricted to the detailed-balance
condition, which constraints the terms in the potential part of
the action. Since we desire to examine the general features of
Ho\v{r}ava gravity, and thus avoiding possible accidental
artifacts of this condition, in this subsection we extend the
investigation beyond detailed balance.

As a demonstration, and without loss of generality, we consider a
detailed-balance-breaking term of the form $\nabla _i R_{jk}
\nabla ^i R^{jk} $. This term induces in the action a second order
term which reads:
\begin{equation}
\nabla _i \delta R_{jk} \nabla ^i \delta R^{jk} = \frac{1}
{4}\left( {\partial _i \nabla ^2 h_{jk} } \right)^2  + \left(
{\partial _i
\partial _j
\partial _k \psi } \right)^2  + 5\left( {\partial _i \nabla ^2
\psi } \right)^2.
 \end{equation}
Thus, the corresponding contribution to the action will be
\begin{equation}
 \delta S^{\left( 2 \right)} _{new}  = \eta \int
{dtd^3 x\left( { - \frac{1} {4}h_{ij} \nabla ^6 h^{ij}  - 6\psi
\nabla ^6 \psi } \right)},
\end{equation}
where $\eta$ is an additional parameter. It is straightforward to
calculate the modifications that $S^{\left( 2 \right)} _{new}$
brings to the dispersion relations for scalar and tensor
perturbations obtained above (expressions (\ref{omegascalar}) and
(\ref{omegatensor}) respectively).
 The extended
dispersion relations read:
\begin{itemize}
\item Scalar perturbations (UV-behavior):
\begin{equation}
\label{scalarnodb} \omega ^2  \sim \frac{{\kappa ^2 \left( {1 -
\lambda } \right)^2 }} {{16\left( {1 - 3\lambda } \right)^2
}}\,k^4 - \frac{{3\kappa ^2 \left( {1 - \lambda } \right)}}
{{2\left( {1 - 3\lambda } \right)}}\eta k^6
\end{equation}
 \item Tensor
perturbations:
\begin{equation}
\label{tensornodb} \omega ^2  = c^2 k^2  + \frac{{\kappa ^4 \mu ^2
}} {{16}}k^4  \pm \frac{{\kappa ^4 \mu }} {{4w^2 }}k^5  + \left(
\frac{{\kappa ^4 }} {{4w^4 }} - \frac{\kappa^{2}\eta}{2}\right)
k^6.
\end{equation}
\end{itemize}
As was expected, the new, detailed-balanced-breaking term,
modifies mainly the UV regime of the theory.

\section{Instabilities, fine-tunings and super-luminal propagation}
\label{discussion}

In the previous section we investigated the gravitational
perturbations of Ho\v{r}ava gravity. Such a study is crucial in
order to examine the consistency of the theory itself. In the
present section we analyze the obtained results, and focus on the
problematic features.

Let us first discuss about the scalar perturbations. As was
mentioned above, (\ref{psi1}),(\ref{psi2}) imply that the regime
$\frac{1}{3} < \lambda < 1$ leads to instabilities. Unfortunately,
this is exactly the flow-interval of $\lambda$-parameter between
the UV and IR regimes. This unstable behavior of the scalar mode
at $\frac{1}{3} < \lambda < 1$, has been already observed in the
literature. In \cite{Lu:2009em} it was argued that it could be
amended by imposing an analytic continuation of the form
\begin{eqnarray}\label{continuation}
\mu\rightarrow i\mu~,~~w^2\rightarrow -iw^2.
\end{eqnarray}
This could also allow for a positive cosmological constant
solution. Unfortunately, performing the analytic continuation
(\ref{continuation}), we straightforwardly see that the UV
behavior is spoiled (see (\ref{omegascalar})) and thus
instabilities re-emerge at high energies. Finally, as we will see
later on, such an analytic continuation would induce similar
problems in the tensor sector, too.

Without analytic continuation, the only physically interesting
case that remains, allowing for a possible flow towards General
Relativity (at $\lambda=1$) is the regime $\lambda\geq1$ (since
the region $\lambda\leq\frac{1}{3}$ is disconnected). Even in this
case though, we cannot evade the instability coming from the
negative mass term, whose sign is independent of both $\lambda$
and $\Lambda$. IR instabilities persist in this regime as long as
we have a non-vanishing cosmological constant. Variations of the
theory with $\Lambda=0$ thus seem to be favored from such a
viewpoint \cite{Kehagias:2009is}. However, as we will see later
on, the $\Lambda=0$ subclass of the original version of Ho\v{r}ava
gravity leads to phenomenological problems concerning the
light-speed definition.

We now turn to the tensor sector. Here again, from
(\ref{omegatensor}) we see that if we desire a well-behaved UV
regime we cannot impose the analytic continuation
(\ref{continuation}). Thus, consistency with the scalar-sector
results means restriction to the $\lambda\geq1$ regime.

Let us now proceed relaxing the detailed-balance condition, that
is analyzing the results of subsection \ref{nodetbal}, according
to which the behavior of detailed-balance Ho\v{r}ava gravity is
modified mainly at the UV. A first and crucial observation is that
the ghost instability of the scalar mode arises from the kinetic
term of the action and thus the breaking of detailed balance,
which affects the potential term, will not alter the
aforementioned scalar-instabilities results concerning the
exclusion of $\frac{1}{3} < \lambda < 1$ regime. In the physical
$\lambda\geq1$ region, the scalar dispersion relation
(\ref{scalarnodb}) remains well-behaved in the UV, provided $\eta$
is negative. Additionally, for negative $\eta$ the UV behavior of
the tensor perturbations is not affected.  Finally, it is
interesting to notice that if $\eta$ is sufficiently negative,
then analytic continuation would not bring problems to the tensor
sector (as can be seen imposing (\ref{continuation}) in
(\ref{tensornodb})). Unfortunately, since analytic continuation
cannot cure the instabilities of the scalar-sector, this
possibility does not have any further physical utility. However,
it can offer an additional indication that a consistent and
well-behaved Ho\v{r}ava gravity should be sought beyond detailed
balance.

In summary, we see that the combined scalar and tensor
perturbation analysis excludes the regime $\frac{1}{3} < \lambda <
1$, since the involved ghost instabilities cannot be removed. Thus, we
conclude that the allowed and physically interesting interval of
Ho\v{r}ava gravity is that with $\lambda\geq1$, and even in this case, the problem of IR instabilities persists. However, it seems
unlikely that a RG flow in this regime will render the theory
power-counting renormalizable, thus negating its initial
motivation as a possible UV completion of gravity. We mention that
this result is valid independently of the imposition of the
detailed-balance condition, and this is the reason we did not
perform an investigation beyond detailed balance in full
generality, examining more such terms. It is also  complementary
to already known works, concerning the presence of the additional
scalar modes and the way they may defer the theory from becoming
equivalent to GR at the IR
\cite{Charmousis:2009tc,Blas:2009yd,Myung:2009sa,Kobakhidze:2009zr}.
This approach is however less ambiguous, since in the regime we
are discussing the new degree of freedom is definitely present.
There is also no need to invoke strong coupling arguments, i.e.
bring matter into the picture \cite{Charmousis:2009tc}.

Let us now discuss a different, but equally significant problem,
that is related to some basic observable consequences of
Ho\v{r}ava gravity. For simplicity we restrict ourselves to the
detailed-balance, but this discussion is independent of that. In
expression (\ref{speedtensor}) we provide the effective light
speed, comparing to the standard light speed $c^2$ given by
(\ref{lightspeed}). If we desire the correction term to be of
next-to-leading order, then $\Lambda$ must be large in general.
However, as it is known the effective cosmological constant in a
universe governed by Ho\v{r}ava gravity is
\cite{Calcagni:2009ar,Kiritsis:2009sh}:
\begin{equation}
|\Lambda_{eff}|=\frac{\kappa^4\mu^2}{16(1-3\lambda)^2}\Lambda^2=\frac{c^2}{|1-3\lambda|}|\Lambda|,
\end{equation}
where the second expression is obtained using (\ref{lightspeed}).
Thus, in general $\Lambda$ should be very small. The
aforementioned contradiction could be resolved by the addition of
a suitable positive constant in an introduced matter sector.
However, having in mind the physical values of the aforementioned
quantities, this would lead to an incredible fine-tuning problem.
In other words, the cosmological constant problem does remain in
Ho\v{r}ava gravity, which is strange since this theory is
constructed to incorporate the underlying and fundamental
gravitational features. Furthermore, the fact that its solution
would demand the contribution of the matter-sector is even more
problematic. Finally, we mention that the aforementioned
light-speed definition makes the $\Lambda=0$ theory problematic
and this was already mentioned in
\cite{Calcagni:2009ar,Kiritsis:2009sh}. Therefore, one cannot
easily set $\Lambda=0$ in order to cure the instabilities as
described above, and thus the aforementioned instability results
seem to be robust.

We close this section by referring to another potential problem.
According to (\ref{speedtensor}), in the physical case
$\lambda\geq1$ the effective light-speed is sub-luminal if
$\Lambda>0$, while it is super-luminal if $\Lambda<0$. However, in
the regime $\lambda\geq1$, $\Lambda$ must be indeed negative in
order to assure for a well-defined light speed
\cite{Calcagni:2009ar,Kiritsis:2009sh}. Therefore, we observe a
possible causality violation. At first it could be stated that
such a violation is not a surprise, since Ho\v{r}ava gravity
violates relativity and thus the light cone is not a bound
anymore. However, it is the classical causality that is violated,
and therefore, one could still construct extensions where both
energy condition and causality would be restored. For such
extensions one could use the formalism and ideas of reggeon field
theory \cite{Abarbanel:1975me,Moshe:1977fe} and of holographic
correspondence used in studies of AdS/CFT
\cite{Janik:2001sc,deBoer:2008gu,Son:2009vu,Giecold:2009cg}. It
seems that Ho\v{r}ava gravity could correspond to a subclass of
models where non-relativistic (super)-conformality, required in
AdS/CFT \cite{Nakayama:2008qm}, can be acquired. Definitely,
causality in Ho\v{r}ava gravity is a subject that requires further
and thorough investigation.

\section{Conclusions}
\label{Conclusions}

In this work we have investigated the gravitational perturbations
of Ho\v{r}ava gravity, with and without the detailed-balance
condition. Setting the relevant degrees of freedom and fixing the
gauge we have extracted the scalar and tensor dispersion relations
and the corresponding effective speeds of propagation. As we have
seen, the scalar sector is plagued by instabilities in the
$\frac{1}{3} < \lambda < 1$ regime, which cannot be cured by
analytic continuation. Thus, the only physical regime, that
additionally allows for an IR limit towards General Relativity, is
the  $\lambda\geq1$ one. This result is in line with previous
treatments discussing potential problems in recovering General
Relativity as an IR limit of the theory. We stress though that in
our case there is no ambiguity about the role of the additional
scalar degree of freedom. Since we are dealing with $\lambda$ away
from unity, the scalar mode is certainly present and no ambiguity
arises. Regardless of possible strong coupling issues occurring at
$\lambda=1$, we see that perturbative instabilities arise long
before we reach the IR limit. The ability of Ho\v{r}ava  gravity
to fulfill its RG flow between the UV and IR, seems to be
jeopardized.

This $\lambda$-restriction also leads to a significant fine-tuning
problem, which is related to the value of the effective
cosmological constant in the universe. This fact casts additional
doubt on Ho\v{r}ava gravity, since any theory that desires to
incorporate some of the fundamental gravitational features, should
provide some route to alleviate the cosmological constant problem.

Additionally, in the present form of Ho\v{r}ava gravity, one could
have causality violation, which would be either a signal of
problematic behavior, or a novel property that originates from the
relativity abandonment. Clearly, a detailed examination of this
subject is crucial in revealing the underlying structure of the
theory.

Although the treatment of the present work is robust at the usual
perturbative framework, we would like to close by commenting on
the possible limitations of our procedure. The presence of a
ghost-like scalar mode may be rendered ambiguous, unless the
entire set of perturbations of the metric is taken into account.
In particular, if there is mixing between different types of
perturbations (that is if the ``orthogonality'' assumption that
forbids the mixing between scalar and tensor modes is not valid),
it may lead to the appearance of ``fictitious'' scalar ghosts,
such as the conformal ghost in General Relativity (see also
\cite{Charmousis:2007ji} and \cite{Garriga:1999yh}). Furthermore,
it could be important to include backreactions of the higher
spatial curvature terms to the geometry, since the presence of
non-linear corrections may lead to non-trivial modifications, as
was recently pointed out in \cite{Mukohyama:2009tp}. Both these
non-trivial extensions could provide additional information on
Ho\v{r}ava gravity and deserve further inquiry.

\vskip .2in \noindent {\large{{\bf {Acknowledgments}}}}

We would like to thank Y.F. Cai, C. Charmousis, S. Mukohyama and
A. Padilla for useful discussions. We also thank K. Sfetsos for
fruitful comments on the earlier version of this paper. C.B. is
supported by the CNRS and the Universit\'e de Paris-Sud XI.

 \vskip
.2in \noindent {\large{{\bf {Note added}}}}

 After the present work was pre-printed  and
 after its submission to the journal, the subject of possible
 instabilities in Ho\v{r}ava gravity has gained a significant
 attention in the literature. In \cite{Wang:2009yz} the authors
 claimed that a more general breaking of the detailed balance can
 cure the instabilities. However, as we showed in our analysis,
 the problematic behavior is not related to the strength of the detailed-balance breaking
but it seems to be a deeper constituent of the theory, at least in
its basic version (with or without detailed-balance) examined in
the present work. These results were verified by many other
authors, and led to generalizations of the basic Ho\v{r}ava
gravity version in order to cure the instabilities. For instance,
in \cite{Blas:2009qj,Koyama:2009hc,Kiritsis:2009vz} it was argued
that the cause of the instabilities is the projectability
condition itself, and thus a new non-projectable version of
Ho\v{r}ava gravity was formulated. But as it was shown in
\cite{Papazoglou:2009fj,Henneaux:2009zb}, even in such an
extension it is ambiguous if the strong coupling problem can be
avoided (see also \cite{Blas:2009ck}). In the same lines, some
other authors claimed that in order to avoid the instabilities,
one should study suitably modified versions of Ho\v{r}ava gravity
\cite{Chen:2009vu} (see also \cite{Park:2009gf} for a different
approach). The evolution of the literature after the appearance of
our work seems to verify our basic result about the problematic
features of the basic version of Ho\v{r}ava gravity. Clearly, the
subject is still open in extended Ho\v{r}ava-like gravitational
theories and deserves further investigation, but this is the
center of interest of separate works.

\appendix
\section{Perturbations of kinetic and potential parts of Ho\v{r}ava  action}\label{appendix}

\subsection{Perturbation of $S_K$}

In order to obtain the perturbed kinetic and  potential terms that
constitute the Ho\v{r}ava gravitational action, we need to extract
the perturbations of the basic geometrical quantities. Following
the formalism of subsection \ref{gauge}, the perturbed spatial
Christoffel symbols read
\begin{eqnarray}
 \delta \Gamma^i_{jk} &=&\frac{1}{2} \left( \partial_j h^i_k + \partial_k h^i_j - \partial^i h_{jk} \right) -\partial_l \partial_k W^i \nonumber\\
&&-\left( \delta^i_k \partial_j \psi + \delta^i_j \partial_k \psi
- \delta_{jk} \partial^i \psi \right) + \partial_k \partial_j
\partial^i E,
 \end{eqnarray}
and thus the perturbation of the Riemann tensor is  $\delta
R^i_{jkl}=\delta\Gamma^i_{jl,k}-\delta\Gamma^i_{jk,l}$, while that
of the Ricci tensor is
 \begin{equation}
 \delta
R_{ij}=-\frac{1}{2} \nabla^2 h_{ij}+\partial_i \partial_j \psi +
\delta_{ij} \nabla^2 \psi,
 \end{equation}
 with
$\nabla^2\equiv\partial_i
\partial^i$.  Therefore, the
  Ricci-scalar perturbation $\delta R= \delta(g^{ij}R_{ij})=\delta
g^{ij} R^{(0)}_{ij}+g^{ij(0)} \delta R_{ij}$ (since
$R^{(0)}_{ij}=0$ and $g^{ij(0)}=\delta^{ij}$) writes as
\begin{equation}
\label{deltaR}
 \delta R = 4 \nabla^2 \psi.
  \end{equation}

Since the extrinsic curvature of the background spacetime is zero,
the only second-order contributions we may have will be of the
form $\delta K_{ij} \delta K^{ij}$ and $(\delta K)^2$. Higher
order contributions arising from the $\sqrt{g}N$ will not enter
here and this coefficient evaluated at zeroth-order is equal to
one. We thus calculate
\begin{eqnarray}
\delta K_{ij} &=& \frac{1}{2} \left( \delta \dot g_{ij} - \partial_i \delta N_j - \partial_j \delta N_i \right)\nonumber\\
&=&\frac{1}{2} \left( \dot h_{ij} - 2 \dot \psi \delta_{ij}
+2\partial_i \partial_j \dot E - \partial_i \dot W_j -\partial_j
\dot W_i \right).
 \end{eqnarray}
Therefore, we also determine $\delta K =
\delta(g^{ij}K_{ij})=\delta g^{ij} K^{(0)}_{ij}+g^{ij(0)} \delta
K_{ij} = \delta^{ij} \delta K_{ij}$, that is
 \begin{equation}
  \delta K =
-3 \dot \psi + \nabla^2 \dot E.
\end{equation}

Assembling everything, the perturbed kinetic part of the action,
up to second order, writes
\begin{equation}
 \delta S^{(2)}_K=\int dt d^3x
\frac{2}{\kappa^2}\left[\frac{1}{4} \dot h_{ij} \dot h^{ij} +
\frac{1}{2} \partial_i \dot W_j \partial^i \dot W^j +(1-3 \lambda)
\left( 3 \dot \psi^2 - 2 \dot \psi \nabla^2 \dot E \right) +
(1-\lambda) \dot E \nabla^4 \dot E \right],
 \end{equation}
  where we
have also used the transversality of fields and integration by
parts to eliminate irrelevant terms and remove derivatives. Thus,
switching off the $W^i$ mode, and concentrating only on scalar and
tensor perturbations, the perturbed kinetic part of the action
(\ref{kinetic}), up to second order, yields the expression
(\ref{DSK}).

\subsection{Perturbation of $S_V$}

Let us now examine the perturbation of potential term
(\ref{potential}). Note that again all quadratic terms must yield
only contributions of the form $\delta R \delta R$ etc, since
these tensors vanish for a flat background. The only terms that do
not follow this rule are the fifth and sixth. The fifth is first
order in $R$, so the second-order perturbation will be of the form
$\delta(\sqrt{g}) \delta R$, while the sixth only receives
contributions at second order from the determinant $\sqrt{g}$. We
will calculate each quadratic expression in turn.

For the first term we need to evaluate the perturbed Cotton tensor
\begin{eqnarray}
\delta C^{ij} &=& \frac{\epsilon^{ikl}}{\sqrt{g^{(0)}}}\partial_k \left( \delta R^j_l -\frac{1}{4} \delta R \delta^j_l \right)\nonumber\\
&=&-\frac{1}{2} \epsilon^{ikl}\partial_k \nabla^2h^j_l,
\end{eqnarray}
which allows to write for the first four-terms:
\begin{eqnarray}
&&\delta C_{ij} \delta C^{ij} = -\frac{1}{4} h_{ij} \nabla^6 h^{ij}\\
&&\epsilon^{ijk} \delta R_{il} \partial_j \delta R^l_k = \frac{1}{4} \epsilon^{ijk}\nabla^2 h_{il} \partial_j \nabla^2 h^l_k\\
&&\delta R_{ij} \delta R^{ij} = \frac{1}{4} \left( \nabla^2 h_{ij}
\right)^2 +6 \psi \nabla^4 \psi,
\end{eqnarray}
together with the square of (\ref{deltaR}). Concerning the
perturbation of the determinant part, we write:
\begin{eqnarray}
\sqrt g NR &=& N^{\left( 0 \right)} \left( {\sqrt {g^{\left( 0
\right)} }  + \frac{1}{{2\sqrt {g^{\left( 0 \right)} } }}\delta g
-
\frac{1}{8}\frac{1}{{\sqrt {g^{\left( 0 \right)} } ^3 }}\delta g^2  +  \ldots } \right)\\
&&\;\;\;\;\;\; \times \left( {\delta ^{ij} R^{\left( 1 \right)}
_{ij}  + \delta ^{ij} R^{\left( 2 \right)} _{ij}
 - \delta g^{ij} R^{\left( 1 \right)} _{ij} } \right) \\
 &=& \delta ^{ij} R^{\left( 1 \right)} _{ij}  + \delta ^{ij} R^{\left( 2 \right)} _{ij}  - \delta g^{ij} R^{\left( 1 \right)} _{ij}
  + \frac{1}{2}\delta g \delta^{ij} R^{\left( 1 \right)} _{ij},
\end{eqnarray}
since the perturbation of $N$ is $\delta N = 0$ and $N^{(0)}=1$.
Note that the combination $\delta ^{ij} R^{\left( 1 \right)}
_{ij}$ is just $\delta R$, and thus the second-order perturbation
becomes
\begin{equation}
\left( {\sqrt g NR} \right)^{\left( 2 \right)}  = \delta ^{ij}
R^{\left( 2 \right)} _{ij}  - \delta g^{ij} R^{\left( 1 \right)}
_{ij}  + \frac{1}{2}\delta g\delta R \label{fifth}
\end{equation}
where, to first order, $\delta g = g^{(0)} g^{ij(0)}\delta
g_{ij}=\delta^{ij} \delta g_{ij} = -6 \psi+2 \nabla^2 E$. Thus the
third term in (\ref{fifth}) becomes
 \begin{equation}
\label{thirdterm}
 \frac{1}{2} \delta g \delta R
= -12 \psi \nabla^2 \psi + 4 \nabla^2 E \nabla^2 \psi.
\end{equation}
Additionally, using integration by parts and imposing
transversality, we obtain
\begin{equation}
\label{secondterm}
 \delta g^{ij} R^{\left( 1 \right)} _{ij}  =  -
\frac{1}{2}h_{ij} \nabla ^2 h^{ij}  - 8\psi \nabla ^2 \psi  +
4\psi \nabla ^4 E.
\end{equation}
Finally, writing the Ricci tensor as
 \begin{equation}
 R^{\left( 2
\right)} _{ij} = \Gamma ^{\left( 2 \right)l} _{ij,l} - \Gamma
^{\left( 2 \right)l} _{il,l}  + \Gamma ^{\left( 1 \right)l} _{lk}
\Gamma ^{\left( 1 \right)k} _{ij}  - \Gamma ^{\left( 1 \right)l}
_{jk} \Gamma ^{\left( 1 \right)k} _{il}
\end{equation}
 where
 \begin{eqnarray}
\Gamma ^{\left( 1 \right)i} _{jk} &=& \frac{1}{2}\left( {\partial _j \delta g^i _k  + \partial _k \delta g^i _j  - \partial ^i \delta g_{jk} } \right)\\
\Gamma ^{\left( 2 \right)i} _{jk}  &=&  - \frac{1}{2}\delta g^{il}
\left( {\partial _j \delta g_{lk}  + \partial _k \delta g_{jl}  -
\partial _l \delta g_{jk} } \right)
 \end{eqnarray}
 with
\begin{eqnarray}
\delta ^{ij} \Gamma ^{\left( 1 \right)l} _{lk} \Gamma ^{\left( 1 \right)k} _{ij}  &=& 3\psi \nabla ^2 \psi  + 2\psi \nabla ^4 E - E\nabla ^6 E\\
\delta ^{ij} \Gamma ^{\left( 1 \right)l} _{jk} \Gamma ^{\left( 1
\right)k} _{il}  &=& \frac{1}{4}h_{ij} \nabla ^2 h^{ij}  + \psi
\nabla ^2 \psi  + 2\psi \nabla ^4 E - E\nabla ^6 E,
 \end{eqnarray}
we conclude that the Ricci scalar can be expressed as
\begin{eqnarray}
 \delta ^{ij} R^{\left( 2 \right)} _{ij}  &=& \delta
^{ij} \Gamma ^{\left( 2 \right)l} _{ij,l}  - \delta ^{ij} \Gamma
^{\left( 2 \right)l} _{il,l}  + \delta ^{ij} \Gamma ^{\left( 1
\right)l} _{lk} \Gamma ^{\left( 1 \right)k} _{ij}  - \delta ^{ij}
\Gamma ^{\left( 1 \right)l} _{jk} \Gamma ^{\left( 1 \right)k}
_{il}\nonumber\\
&=&  - \frac{1}{4}h_{ij} \nabla ^2 h^{ij}  + 2\psi \nabla ^2 \psi.
 \label{Ricci2}
 \end{eqnarray}
Assembling (\ref{thirdterm}),(\ref{secondterm}),(\ref{Ricci2}) we
find that  the fifth term in the perturbed action reduces to
\begin{equation}
\left( {\sqrt g NR} \right)^{\left( 2 \right)}  =
\frac{1}{4}h_{ij} \nabla ^2 h^{ij} - 2\psi \nabla ^2 \psi.
\end{equation}

Lastly, using $\delta \sqrt{g} = \frac{1}{2 \sqrt{g^{(0)}}} \delta
g = -3 \psi + \nabla^2 E$ and $(\delta
\sqrt{g})^{(2)}=-\frac{1}{8} \delta g^2 = -\frac{9}{2} \psi^2 +3
\psi \nabla^2 E -\frac{1}{2} (\nabla^2 E)^2$, the sixth term in
the perturbed action reads
\begin{equation}
 (\delta \sqrt{g})^{(2)} N^{(0)} = -\frac{9}{2}
\psi^2 +3 \psi \nabla^2 E -\frac{1}{2} (\nabla^2 E)^2.
\end{equation}

Finally, assembling the aforementioned terms, we conclude that the
perturbed potential part of the action is (\ref{DSV}).

\end{document}